\documentclass{emulateapj}
\usepackage{graphicx}
\begin{document}

\def\ba{\begin{eqnarray}}
\def\ea{\end{eqnarray}}
\def\etal{et al.\ \rm}

\title{Stellar proper motion and the timing of planetary transits}

\author{Roman R. Rafikov\altaffilmark{1,2}}
\altaffiltext{1}{Department of Astrophysical Sciences, 
Princeton University, Ivy Lane, Princeton, NJ 08540; 
rrr@astro.princeton.edu}
\altaffiltext{2}{Sloan Fellow}


\begin{abstract}
Duration and period of transits in extrasolar planetary systems 
can exhibit long-term variations for a variety of reasons. 
Here we investigate how systemic proper motion, which steadily 
re-orients planetary orbit with respect to our line of sight, affects 
the timing of transits. We find that in a typical system with a 
period of several days proper motion at the level of 100 mas yr$^{-1}$ 
makes transit duration vary at a rate $\sim 10-100$ ms 
yr$^{-1}$. In some isolated systems this variation is at 
the measurable level (can be as high as $0.6$ 
s yr$^{-1}$ for GJ436) and may exceed all other transit timing 
contributions (due to the general relativity, stellar quadrupole, etc.). 
In addition, proper motion causes evolution of the observed 
orbital period via the Shklovskii effect at a rate  
$\gtrsim 10$ $\mu$s yr$^{-1}$ for the nearby transiting systems 
($0.26$ ms yr$^{-1}$ in GJ436), which in some cases 
exceeds all other contributions to $\dot P$. Earth's motion
around the Sun gives rise to additional periodic timing signal
(even for systems with zero intrinsic proper motion) allowing 
a full determination of the spatial orientation of the planetary 
orbit. 
Unlike most other timing effects the proper motion signatures 
persist even in systems with zero eccentricity and get 
stronger as the planetary period increases. They should be 
the dominant cause of transit timing 
variations in isolated wide separation (periods 
of months) systems that will be sought by {\it Kepler}.
\end{abstract}
\keywords{astrometry --- celestial mechanics --- planetary systems --- 
eclipses}


\section{Introduction.}
\label{sect:intro0}


Planetary transits 
have provided us with a unique opportunity to get a handle on the 
physical properties of the extrasolar planets such as their radii 
and densities. Recently it has 
been suggested (Miralda-Escud\'e 2002; Heyl \& Gladman 2007) 
that precision timing of the moments at which transits occur 
can give us additional information about the transiting systems. 
Various physical effects cause orbit of the planet
precess in space leading to the changes in transit geometry, which  
can be measured through the timing of transits. Among these effects 
are the general
relativistic precession of the orbit, gravitational influence of other planets 
in the system or companion stars, torques due to the spin-induced 
quadrupole moment of the star and due to the tidal deformations 
of both the star and the planet (Miralda-Escud\'e 2002; 
Heyl \& Gladman 2007; Ribas \etal 2008; P\'al \& Kocsis 2008;
Jord\'an \& Bakos 2008). 

Another obvious reason for the re-orientation of the planetary
orbit with respect to observer at Earth is the proper motion of the 
exoplanetary system with respect to the Solar System. Stars in the Solar 
neighborhood move at velocities of tens of km s$^{-1}$ and some 
of them exhibit proper motion at the level of $1$ mas 
yr$^{-1}$. Also, the distance to stars constantly changes as a result of 
their relative motion with respect to the Sun and this affects 
transit timing because of the finite speed of light. At some level 
proper motion is a characteristic of any star, including those with 
transiting exoplanets, and it is thus important to understand its 
implications for transit timing.

Proper motion is well known to be important in the timing of 
isolated and binary radio pulsars (Shklovskii 1970; 
Kopeikin 1996). In these systems proper motion affects the pulsar spin
and orbital periods through the so-called Shklovskii effect (Shklovskii 1970) 
while the re-orientation of the binary orbit can be (and has been) measured 
via the variation of the projected size of the orbit (Kopeikin 1996; Arzoumanian 
\etal 1996). Pulsar acceleration in external gravitational field can also 
be important especially for pulsars in globular clusters 
(Edwards \etal 2006). 

Of course, there are significant differences between the timing of pulsars
and of planetary transits: accuracy with which some millisecond pulsars 
can be timed is at the $\mu$s level (Manchester 2008) while a single
planetary transit can only be timed to several seconds at 
best (Knutson \etal 2007). 
Also, the whole idea of timing is different in the two cases: for binary 
pulsars one is usually able to trace the whole orbit of 
the neutron star in time domain while in the case of planetary 
transits only two narrow time windows --- primary and secondary 
transits --- are available to play with. Nevertheless, some of the 
ideas developed in pulsar timing may be applied to the timing of 
planetary transits. 

Previously, Kopeikin 
\& Ozernoy (1999) have utilized a post-newtonian relativistic 
approach for the precision Doppler measurements of the binary star 
orbits and discussed some of the relevant effects of the proper motion. Here 
we aim at investigating the role of the proper motion in timing of 
planetary transit duration and period in extrasolar planetary systems.
We lay out the basics of the orbital element evolution
due to the proper motion in \S \ref{sect:intro} . We discuss the evolution
of the transit duration in \S \ref{sect:transit} and the evolution of
the period between transits in \S \ref{sect:period}. Comparison with
other transit timing effects and application to real systems can be 
found in \S \ref{sect:disc}.


\section{Effect of proper motion.}
\label{sect:intro}


To quantitatively evaluate the effect of stellar proper motion
on the timing of planetary transits let us consider a planet
in orbit around a star with period $P$, semimajor axis $a$,
and eccentricity $e$. We introduce a unit
vector ${\bf n}$ pointing from the observer at Earth to 
the barycenter of the transiting system. Vector ${\bf n}$
varies in time because of the linear motion of the binary:
\ba
\frac{d{\bf n}}{dt}={\bf \mu},
\label{eq:dndt}
\ea
where ${\bf \mu}$ is the proper motion in the plane of the sky.
Orientation of
the binary in space is fully determined by the unit vector
${\bf l}$ parallel to the orbital angular momentum 
${\bf L}$ of the binary (i.e. ${\bf l}$ is perpendicular to the orbital 
plane) and the unit vector ${\bf g}$ pointing from the prime
focus of the planetary orbit towards its pericenter. 
We assume ${\bf L}$ to be constant thus neglecting 
possibility of tidal coupling between ${\bf L}$ and planetary
and stellar spins, and gravitational effects of
any companions. We also assume that orientation of the orbital 
ellipse in space is fixed, i.e. ${\bf g}$ is constant too. 
In doing this we disregard precession of 
the planetary orbit caused by the general relativity, stellar
oblateness, and so on. We can do this because 
observed changes of the orbital configuration caused by 
different physical mechanisms add up linearly and here
we want to concentrate on just one of them.

Orbital plane crosses the plane of the sky 
along the line of nodes and we introduce vector 
${\bf m}=({\bf l}\times{\bf n})/\sin i$ along this line 
($|{\bf m}|=1$), where $i$ is the observed inclination 
of the planetary orbit given by $\sin i=|{\bf n}\times{\bf l}|$. 
If $\omega$ is the angle 
between ${\bf m}$ and ${\bf g}$ in the direction of planetary
motion --- the argument of pericentre --- then at any moment of time 
\ba
{\bf g}=\frac{\cos\omega}{\sin i}({\bf l}\times{\bf n})-
\frac{\sin\omega}{\sin i}[{\bf n}-{\bf l}({\bf n}\cdot{\bf l})].
\label{eq:p_vector}
\ea 

Differentiating relation $\cos i=({\bf n}\cdot{\bf l})$ with
respect to time we find using equation (\ref{eq:dndt})
\ba
\dot i_\mu=-\frac{({\bf \mu}\cdot{\bf l})}{\sin i}=
-\mu\cos\beta,
\label{eq:didt}
\ea
where $\mu=|{\bf \mu}|$ and $\beta$ is the angle in the plane 
of the sky between ${\bf \mu}$ and vector 
${\bf l}-{\bf n}({\bf l}\cdot{\bf n})$ --- the projection of ${\bf l}$
on the sky plane. 
Differentiating with respect to time relation
$\cos \omega = ({\bf g}\cdot{\bf m})=({\bf g}
\cdot({\bf l}\times{\bf n}))/\sin i$ and using equations
(\ref{eq:dndt}), (\ref{eq:p_vector}), and (\ref{eq:didt}) we 
find (Kopeikin 1996)
\ba
\dot \omega_\mu = -\frac{({\bf \mu}\cdot({\bf l}\times {\bf n}))}
{\sin^2 i}=-\frac{\mu\sin\beta}{\sin i}.
\label{eq:domegadt}
\ea
Equations (\ref{eq:didt}) and (\ref{eq:domegadt}) fully determine 
evolution of the observed orientation of planetary orbit in space 
caused by the stellar proper motion.


\section{Variation of the transit duration.}
\label{sect:transit}


Planet transit is characterized by an impact parameter 
$p=r_{tr}\cos i/R_\star$ --- minimum separation
between the planetary trajectory and the stellar disk center projected 
onto the plane of the sky, in units of stellar radius $R_\star$. 
Here $r_{tr}$ is the value of the 
spatial separation $r$ between the planet and the center of the star 
at transit midpoint --- moment of time when the projected separation 
between the planet and the center of the stellar disk is minimized.  
In general 
\ba
r=\frac{a(1-e^2)}{1+e\cos f},
\label{eq:r}
\ea
where $f$ is the true anomaly counted from the line of apsides.
Transit midpoint occurs at $f=\pi/2-\omega$, so that
\ba
p=\frac{a}{R_\star}\frac{\cos i(1-e^2)}{1+e\sin \omega}.
\label{eq:p}
\ea 
Clearly, for the transit to occur one needs $p<1+R_p/R_\star$, where 
$R_p$ is the planetary radius, which translates
into
\ba
\cos i < \frac{R_\star+R_p}{a}\frac{1+e\sin \omega}{(1-e^2)}.
\label{eq:constr}
\ea

Transit duration calculated as the time between the crossings of the
edge of the stellar disk by the center\footnote{This definition differs
from that usually adopted in the literature (which assumes that transit 
lasts while stellar and planetary disks have at least some overlap) but this
does not affect our results significantly.} of the planetary disk is
(see e.g. Tingley \& Sackett 2005)
\ba
T_{tr}=2\frac{R_\star(1-p^2)^{1/2}}{v_{\varphi,tr}}=
\frac{2}{n}\frac{(1-e^2)^{1/2}}{1+e\sin \omega}
\frac{R_\star(1-p^2)^{1/2}}{a},
\label{eq:T_tr}
\ea
where $v_{\varphi,tr}=na(1+e\sin \omega)/(1-e^2)^{1/2}$ is the value
of the azimuthal (transverse) component of planetary velocity at
the transit midpoint and $n=2\pi/P$ is the planetary mean motion. 
In deriving equation (\ref{eq:T_tr}) we have neglected the curvature 
of projected planetary trajectory and the variation of planetary speed 
during the transit --- this introduces only a  small error. 

Given that $p$ and $\omega$ in equation (\ref{eq:T_tr}) evolve as a 
result of stellar proper motion it is obvious that 
$T_{tr}$ would not remain constant. Differentiating expression 
(\ref{eq:T_tr}) with respect to time  one finds
\ba
&& \dot T_{tr}=-\frac{T_{tr}}{1+e\sin \omega}\nonumber\\
&& \times\left[e\dot\omega\cos\omega-
g\left(
\dot i\sin i+\dot 
\omega\cos i\frac{e\cos\omega}{1+e\sin \omega}\right)\right],
\label{eq:dTtrdt}
\ea
where
\ba
g\left(\frac{R_\star}{a},p\right)
=\frac{a}{R_\star}\frac{p}{1-p^2}.
\label{eq:g}
\ea
In equation (\ref{eq:dTtrdt}) the first term in brackets describes the
variation of $T_{tr}$ caused by the change of $v_{\varphi,tr}$ due to
the precession of the orbital ellipse while the second and the third terms 
embody the variation of transit geometry (change of impact parameter $p$) 
caused by the change of the inclination of the orbital plane and the precession 
of the orbital ellipse respectively. 
Third term is normally much smaller than the second one because
$\cos i\ll 1$ in transiting systems. 
Note that $\dot \omega$ affects $T_{tr}$ only if the planetary
orbit is eccentric, while $\dot i$ causes variation of $T_{tr}$
even for circular orbits.

Expression for $\dot T_{tr}$ caused by the proper motion 
can be written with the aid of equations (\ref{eq:didt}), 
(\ref{eq:domegadt}), and (\ref{eq:dTtrdt}) as
\ba
&& \dot T_{tr,\mu}=\frac{T_{tr}\mu\sin\beta}{1+e\sin \omega}\nonumber\\
&& \times\left[\frac{e\cos\omega}{\sin i}-
g\left(
\frac{\sin i}{\tan\beta}+\frac{\cos i}{\sin i}\frac{e\cos\omega}
{1+e\sin \omega}\right)\right].
\label{eq:dTtrdt1}
\ea
This equation explicitly shows how $T_{tr}$ varies as a function of 
the absolute value of the stellar proper motion $\mu$ and the 
orientation of ${\bf \mu}$ with respect to the 
projection of the orbital angular momentum onto the plane of 
the sky --- angle $\beta$.


\section{Variation of the orbital period.}
\label{sect:period}


Precession of planetary orbit makes observed orbital period
of the planet $P_{obs}$ different from the true orbital
period $P$. Indeed, suppose that we try to determine the 
period of transiting planet by measuring the time between 
the successive inferior conjunctions of the planet. Every 
orbital period precession at a uniform rate $\dot\omega$ 
turns the orbit by an angle
$\Delta \varphi=P\dot\omega$ with respect to our line of 
sight which gets reflected in the length of the time interval 
between successive conjunctions. The extra time it takes a planet to
cover this additional angle is 
$\Delta P_\omega=\Delta\varphi/\dot\varphi$, where $\dot\varphi=
v_{\varphi,tr}/r_{tr}$ is the angular frequency of the planet
at the point of conjunction. Using equation (\ref{eq:r}) and
expression for $v_{\varphi,tr}$ we can write the deviation of 
observed planetary period from the true one as
\ba
\frac{\Delta P_\omega}{P}=-\frac{\dot\omega}{n}\frac{(1-e^2)^{3/2}}
{(1+e\sin\omega)^2}.
\label{eq:dPmu}
\ea
Note that $\Delta P_\omega$ is nonzero even in the case of circular 
orbits in agreement\footnote{By contrast expressions for 
$\Delta P_\omega$ derived in Miralda-Escud\'e (2002) and 
P\'al \& Kocsis (2008) vanish in the limit $e\to 0$.} 
with Kopeikin (1996) and  Heyl \& Gladman (2007). 
In general $\dot\omega$ in equation (\ref{eq:dPmu}) is
given by
\ba
\dot\omega=\dot\omega_\mu+\dot\omega_{GR}+\dot\omega_S,
\label{eq:dot_omega}
\ea
where different terms on the right hand side represent apparent orbital 
precession caused by the systemic proper motion, general relativity  
and the quadrupole moment of the central star correspondingly (other 
sources of orbital precession, e.g. tidal deformations of star and 
planet (Jord\'an \& Bakos 2008) have been 
neglected here for simplicity). As a result,
$\Delta P_\omega=\Delta P_{\omega,\mu}+\Delta P_{\omega,GR}+
\Delta P_{\omega,S}$, where $\Delta P_{\omega,\mu}$, $\Delta P_{\omega,GR}$
and $\Delta P_{\omega,S}$ are found by substituting $\dot\omega_\mu$, 
$\dot\omega_{GR}$, and $\dot\omega_S$ for $\dot\omega$ in equation 
(\ref{eq:dPmu}).

Apart from $\Delta P_\omega$ which owes its existence to the 
apparent {\it re-orientation} of the planetary orbit there
is another contribution to $P_{obs}$ related to the systemic motion: 
the {\it distance} to the planetary system changes, which because of the finite 
speed of light gives rise to a special relativistic contribution 
$\Delta P_{rel}$ given by 
\ba
\frac{\Delta P_{rel}}{P}=\frac{v_r}{c},
\label{eq:DeltaP_SR}
\ea 
where $v_r$ is the line-of-sight velocity of the system (positive for 
systems moving away from us). Thus, in general 
\ba
P_{obs}=P+\Delta P_\omega+\Delta P_{rel}.
\label{eq:dPobs}
\ea

One cannot, of course, measure $\Delta P$ directly since the true orbital 
period of the system is not known a priori. However, one might try to 
measure the {\it variation} of $P$ over an extended period of time. With 
equations (\ref{eq:dPmu}), (\ref{eq:DeltaP_SR}), and (\ref{eq:dPobs}) 
one can easily show that
\ba
\dot P=\dot P_\omega+\dot P_{Shk},
\label{eq:dotP}
\ea
where
\ba
&& \dot P_\omega=-\frac{2\pi}{n^2}
\frac{(1-e^2)^{3/2}}{(1+e\sin \omega)^2}
\left[\ddot\omega-2(\dot\omega)^2\frac{e\cos\omega}
{1+e\sin\omega}\right],
\label{eq:dotPomega}\\
&& \dot P_{Shk}=\dot P_{rel}=\frac{Pv_t^2}{c D}=\frac{P\mu^2 D}{c},
\label{eq:dotPrel}
\ea
with $D$ being the distance to the planetary system and $v_t=\mu D$ 
being its transverse velocity.
The timing contribution $\dot P_{Shk}$, which to the best of our knowledge 
has never been highlighted in the context of planetary transit timing, is 
identical to the so-called Shklovskii effect well known from pulsar timing 
(Shklovskii 1970): radial motion of the system changes the observed orbital 
period via the Doppler effect but the radial component of the velocity (and the 
Doppler factor) varies if there is a non zero transverse component of 
the systemic velocity, leading to non zero $\dot P$. This contribution to 
$\dot P$ is always positive since spatial motion of the planetary system 
always increases $v_r$. In \S \ref{sect:disc} we demonstrate that in 
many transiting systems Shklovskii effect dominates $\dot P$. 

Equation (\ref{eq:dotPomega}) generalizes expressions for $\dot P_\omega$
obtained by Heyl \& Gladman (2007) and P\'al \& Kocsis (2008) by including
the term proportional to $\ddot\omega$. One expects $\ddot\omega\ll
(\dot\omega)^2$ for precession caused by the general relativity and the stellar
quadrupole in which case equation (\ref{eq:dotPomega}) reduces to the
expression derived by other authors. However, in the case of apparent precession
caused by the systemic motion one can easily show using results of 
\S \ref{sect:intro} that $\ddot\omega_\mu\sim
\mu^2\sim (\dot\omega_\mu)^2$, so that all terms in equation (\ref{eq:dotPomega})
for $\dot P_{\omega,\mu}$ must be retained. In general, $\dot P_\omega
\ll\dot T_{tr}$ because $\dot T_{tr}$ is a linear function of the small
parameter $P\dot\omega$ while $\dot P_\omega$ is quadratic.


\section{Discussion.}
\label{sect:disc}


Here we compare the effects caused by the proper motion with
other timing contributions and discuss their observability in
different types of systems. 

As a fiducial system we will take a star located 100 pc away 
from the Sun and moving with transverse velocity $30$ km s$^{-1}$.
Such a system has proper motion $\mu\approx 60$ mas yr$^{-1}$ resulting 
in $\dot i_\mu,\dot\omega_\mu\approx
2\times 10^{-7}$ yr$^{-1}$ for $\beta=45^\circ$ and $i=90^\circ$. Timescale 
on which planetary orbit changes its orientation is $\sim \mu^{-1}\sim 
5\times 10^{6}$ yrs. We can compare $\dot\omega_\mu$ to the general
relativistic periastron precession rate 
\ba
\dot \omega_{GR}& = &\frac{3n}{1-e^2}\left(\frac{na}{c}\right)^2
\label{eq:dotomegaGR}\\
& = &
\frac{3.7\times 10^{-4}}{1-e^2}
\left(\frac{10R_\odot}{a}\right)^{5/2}\mbox{yr}^{-1},\nonumber
\ea
and to the rate of precession due to the rotation-induced 
stellar quadrupole (Miralda-Escud\' e 2002)
\ba
\dot\omega_S\approx n\frac{3 J_2 R_\star^2}{2 a^2}\approx 
9\times 10^{-6}\frac{J_2}{10^{-6}}
\left(\frac{10R_\odot}{a}\right)^{7/2}\mbox{yr}^{-1}
\label{eq:dotomegaS}
\ea
where $J_2$ is the dimensionless measure of the stellar quadrupole moment
(its typical value for the Solar type stars is $J_2\sim 10^{-6}$) and 
we took $M_\star=M_\odot$ and $R_\star=R_\odot$.

These estimates clearly indicate that for Solar type stars with short 
period ($P=3-4$ days) planets $\dot \omega_{GR}\gg\dot \omega_S\gg 
\dot i_\mu,\dot \omega_\mu$. Plugging expression (\ref{eq:dotomegaGR})
into equation (\ref{eq:dotPomega}) we find that a planetary system with 
$M_\star=M_\odot$, $e=0.1$ and $\omega=45^\circ$ should exhibit
\ba
\dot P_{\omega,GR}&=&-\frac{36\pi e\cos\omega}{(1-e^2)^{1/2}
(1+e\sin\omega)^3}\left(\frac{na}{c}\right)^4
\label{eq:dotPomegaGR}\\
&=& 8.8
\left(\frac{10R_\odot}{a}\right)^{2}\mu\mbox{s yr}^{-1}.\nonumber
\ea
Given that $\dot \omega_\mu\ll \dot \omega_{GR}$ it is clear that 
$\dot P_{\omega,\mu}\ll\dot P_{\omega,GR}$ so that 
the re-orientation of the planetary orbit caused by the stellar proper motion does 
not noticeably affect $\dot P_\omega$ (the same is true for the
precession caused by the stellar quadrupole since
$\dot \omega_S\ll\dot \omega_{GR}$). 

However, this does not mean that one can just ignore the effect of the 
proper motion on $\dot P$: proper motion also affects $\dot P$
via the Shklovskii effect and the magnitude of this contribution
\ba
\dot P_{Shk}&=& 9.6\left(\frac{v_t}{30~\mbox{km s}^{-1}}\right)^2
\frac{100~\mbox{pc}}{D}\left(\frac{a}{10R_\odot}\right)^{3/2}
\mu\mbox{s yr}^{-1}\nonumber\\
&=& 20\left(\frac{\mu}{100~\mbox{mas yr}^{-1}}\right)^2
\frac{D}{100~\mbox{pc}}\frac{P}{3~\mbox{d}}
~\mu\mbox{s yr}^{-1}
\label{eq:dotPrel_num}
\ea
may be comparable to $\dot P_{\omega,GR}$. Clearly, $\dot P_{Shk}$ 
can be quite important even for tight, eccentric systems for which
one would normally expect $\dot P_{\omega,GR}$ to dominate.  

One also has to keep in mind that the majority 
of short period transiting systems have eccentricities consistent
with zero. In such systems with circular orbits $\dot P_{\omega,GR}$ and 
$\dot P_{\omega,S}$ vanish (remember that $\ddot \omega_{GR},
\ddot \omega_S\approx 0$) leaving Shklovskii effect as the only 
source of non zero $\dot P_\omega$
at the level of tens of $\mu$s per year. In Table \ref{table} we have 
summarized the properties of observed transiting systems 
(supplemented with two artificial systems Sys-1 and Sys-2 with the goal 
of illustrating transit timing effects in long period systems) in which 
proper motion effects are particularly pronounced 
(namely, max$|\dot T_{tr,\mu}|>10$ 
ms yr$^{-1}$), while in Table \ref{table2} we display the values
of various timing contributions in these systems,
including $\dot P_{\omega,GR}$ and  $\dot P_{Shk}$. From 
Table \ref{table2} one can see that in some nearby 
high-proper motion systems like GJ436 $\dot P_{Shk}$ is a good fraction of 
ms yr$^{-1}$. Such high rate of period change significantly 
exceeds $\dot P_{\omega,GR}$ and may in principle be measurable on 
a time scale of tens of years assuming observing parameters typical for the 
{\it Kepler} photometric mission (Miralda-Escud\'e 2002; Jord\'an \& 
Bakos 2008).

Variation of the transit duration $T_{tr}$ presents another 
way of detecting proper motion effects in isolated star-planet
systems, as described in \S \ref{sect:transit}. Assuming that all angle 
dependent factors in equation (\ref{eq:dTtrdt1}) are of order unity one 
finds
\ba
\dot T_{tr,\mu} & \sim & g T_{tr}\mu
\label{eq:dTtrdt_num}\\
& \approx & 50\frac{\mu}{100~\mbox{mas yr}^{-1}}
\frac{a/R_\star}{10}\frac{T_{tr}}{4~\mbox{hr}}~\mbox{ms yr}^{-1},
\nonumber
\ea
where in evaluating $g$ we have assumed $p=0.5R_\star$. Thus, a typical 
nearby exoplanetary system indeed exhibits $\dot T_{tr,\mu}\gg\dot P$. 
A specific value of $\dot T_{tr,\mu}$ for a particular exoplanetary
system depends not only on $\mu$ but also (sinusoidally) on the angle 
$\beta$ between ${\bf \mu}$ and the line of nodes. The maximum possible 
value of $\dot T_{tr,\mu}$ for several 
representative systems can be found in Table \ref{table2}.

At the same time, for $M_\star=M_\odot$, $e=0.1$ and $\omega=45^\circ$ 
one finds from equation (\ref{eq:dTtrdt}) the following value of 
$\dot T_{tr}$ due  to the general relativity:
\ba
\dot T_{tr,GR}\approx 240\left(\frac{10R_\odot}{a}\right)^{5/2}
\frac{T_{tr}}{4~\mbox{hr}}
~\mbox{ms yr}^{-1}.
\label{eq:dTtrdt_GR}
\ea
This is not much larger than $\dot T_{tr,\mu}$ and in some high proper 
motion systems $\dot T_{tr,\mu}$ may even dominate. The best example 
is GJ436: because system is very compact general relativity 
provides $\dot T_{tr,GR}\approx -0.2$ s yr$^{-1}$ but the very high proper 
motion of the system ($\mu\approx 1.2$ mas yr$^{-1}$) gives rise to 
max$|\dot T_{tr,\mu}|\approx 0.6$ s yr$^{-1}$. 
Thus, in general one cannot simply ascribe all $\dot T_{tr}$ measured in 
eccentric systems to the general relativity --- some fraction of $\dot T_{tr}$ can
also be contributed by the proper motion. In systems with circular orbits 
$\dot T_{tr,GR}=0$. 

Given that $\dot\omega_\mu\ll\dot\omega_{GR}$ the magnitude of the
effect of the proper motion on $\dot T_{tr}$ may seem 
disproportionately large compared to $\dot T_{tr,GR}$. The reason
for this lies in the amplifying 
factor $g$ in equation (\ref{eq:dTtrdt}) which propagates into 
$\dot T_{tr,\mu}$, see equation (\ref{eq:dTtrdt1}). According to the equation
(\ref{eq:p}) the magnitude of $g$
is determined by the transit impact parameter $p$ and the ratio
$a/R_\star$ which is usually quite large, $\sim 10$ even for rather
short period ($P=3-4$ d) systems. For grazing transits (such as those 
occurring in GJ436, see Table \ref{table2}), when $1-p\ll 1$,
$g$ gets additionally boosted up\footnote{Thus, monitoring systems with 
large $p$ tends to increase the chances of measuring $\dot T_{tr}$.}
because then $T_{tr}$ becomes a very sensitive function of $p$ and $i$, 
see Ribas \etal (2008). At the same time factor $g\gg 1$ does not greatly 
affect $\dot T_{tr,GR}$ since for precession induced by the 
general relativity $\dot i=0$ and $g$ enters the expression for $\dot T_{tr,GR}$ 
only in combination $g\cos i$ while $\cos i\ll 1$ in transiting 
systems ($\cos i\lesssim R_\star/a$ so that $g\cos i\sim 1$, see equations
(\ref{eq:p}) and (\ref{eq:g})). This explains why 
$\dot T_{tr,GR}\sim\dot T_{tr,\mu}$ even though $\dot \omega_{GR}\gg
\dot \omega_\mu$.

Inclination of the planetary orbit with respect to our line of sight
may change also because of the spin induced quadrupole if the stellar spin
axis is misaligned with the orbital angular momentum vector. The spin 
induced $\dot T_{tr,S}$ is amplified by factor $g$ in a way 
analogous to the amplification of $\dot T_{tr,\mu}$. Given that in some
systems $\dot\omega_S$ can be 1-2 orders of magnitude larger than 
$\dot\omega_\mu$ (see Table \ref{table2} where $\dot\omega_S$ is computed 
for $J_2=10^{-6}$) one may expect 
$\dot T_{tr,S}\gg\dot T_{tr,\mu}$ in these systems. However, in reality 
it will often be the case that
$\dot i_S\ll\dot\omega_S$ since it can be 
demonstrated that $\dot i_S=C\dot\omega_S\sin\lambda$ (Lai \etal 1995), 
where $C\sim 1$ is the angle-dependent
factor and $\lambda$ is the angle between the stellar spin axis and the
orbital angular momentum vector. Misalignment angle $\lambda$ has been 
measured in several systems via the Rossiter-McLaughlin effect 
(Rossiter 1924; McLaughlin 1924) and in the majority of measured cases
$\lambda$ is close to zero, as expected from the planet formation 
theories. Among the systems in Table \ref{table} for which $\lambda$
has been measured this angle was found to be small in HD189733 
($\lambda=1.4^\circ\pm 1.1^\circ$, Winn \etal 2006) and HAT-P-1
($\lambda=3.7^\circ\pm 2.1^\circ$, Johnson \etal 2008) while 
in HD17156 misalignment may be significant ($\lambda=62^\circ\pm 25^\circ$, 
Narita \etal 2008), although Cochran \etal (2008) have found 
$\lambda=9.4^\circ\pm 9.3^\circ$ in this system. 
In HD189733 $\dot i_S$ end up being $\ll\dot i_\mu$
so that $\dot T_{tr,S}$ likely makes negligible contribution to  
$\dot T_{tr}$ which should be dominated by the proper motion. 
In HAT-P-1 we find $\dot i_S\sim \dot i_\mu$ and $\dot T_{tr,S}\sim\max 
|\dot T_{tr,\mu}|$ with $\dot T_{tr,GR}$ providing a non-negligible contribution. 
Finally, in HD17156, if we adopt a larger value of $\lambda$ found by 
Narita \etal (2008), $\sin\lambda\sim 1$ but $\dot i_S$
is still comparable to $\dot i_\mu$ because the semimajor axis of the system 
is quite large which greatly reduces $\dot\omega_S$. As a result, 
$\dot T_{tr,S}\sim \max|\dot T_{tr,\mu}|$ in this 
system and both are somewhat smaller than $\dot T_{tr,GR}$.
Thus, at least in the systems presented in Table \ref{table} the spin-induced
quadrupole orbital precession does not strongly exceed the proper motion 
effects in timing of transit duration.

Note that the tidal deformations induced on the star and the planet by 
each other affect $\dot T_{tr}$ in a way different from
that of the spin induced quadrupole --- similar to the general relativity 
the tidal bulges do not generate non zero $\dot i$. 
Given that tidal $\dot \omega$ is typically smaller 
than $\dot\omega_{GR}$ (Jord\'an \& Bakos 2008) we may conclude that 
tidally induced $\dot T_{tr}$ is lower than $\dot T_{tr,GR}$ and is thus 
$\lesssim \dot T_{tr,\mu}$. 

It is obvious from the preceding discussion that the proper motion can 
have an appreciable (if not dominant in some cases) effect on transit 
timing in the short period 
systems. This statement becomes much more robust when we go to systems with 
wider separations. It is obvious from equations (\ref{eq:dotomegaGR}), 
(\ref{eq:dotomegaS}), (\ref{eq:dotPomegaGR}), and (\ref{eq:dTtrdt_GR})
that $\dot\omega_{GR}$, $\dot\omega_S$, and all contributions to $\dot P$
and $\dot T_{tr}$ caused by the effects of the general relativity, 
stellar quadrupole and tidal deformations are rapidly decreasing functions 
of $a$. At the same time, $\dot\omega_\mu,\dot i_\mu$ are independent of $a$ while
both $\dot P_{Shk}$ and $\dot T_{tr}$ increase quite rapidly with $a$, 
see equations (\ref{eq:dotPrel_num}) and (\ref{eq:dTtrdt_num}). This means  
that the proper motion should completely dominate transit timing 
variations in isolated (i.e. containing no other planets) wide 
separation systems. For example, a 
transiting planet in a 30 d orbit around a Solar type star would 
exhibit
$\dot P_{\omega,GR}\approx 0.6$ $\mu$s yr$^{-1}$ and 
$\dot T_{tr,GR}\approx 7.5$ ms yr$^{-1}$ if $e=0.1$ and $T_{tr}=4$ hr.
If this system is located 100 pc away from the Sun and has proper motion
$\mu=100$ mas yr$^{-1}$ then one finds $\dot P_{Shk}\approx 200$ 
$\mu$s yr$^{-1}$ and $\dot T_{tr,\mu}\approx 200$ ms yr$^{-1}$, so
that both $\dot P_{\omega,GR}\ll\dot P_{Shk}$ and 
$\dot T_{tr,GR}\ll\dot T_{tr,\mu}$. To additionally illustrate the 
importance of proper motion for wide separation systems we introduce 
two artificial systems (Sys-1 and Sys-2) in Table \ref{table} and calculate 
their timing 
parameters in Table \ref{table2}. Such wide separation systems 
are one of the primary goals of photometric missions like {\it Kepler}.
It is clear that if such systems are found to 
exhibit transit timing variations then these variations 
must be caused by the systemic proper motion, provided that the influence 
of possible additional companions is proven to be negligible. In this case 
according to equation (\ref{eq:dTtrdt1}) the measurement of $\dot T_{tr}$ 
would serve as a measurement of angle $\beta$ giving us information on
the full three dimensional orientation of the transiting system.
However,
one must remember that if additional planets in external orbits are present 
in these systems then their influence may not be disregarded (Ribas \etal 
2008) since
their effect on transit timing grows with $a$ faster than 
$\dot T_{tr,\mu}$ and $\dot P_{Shk}$ do. 

Note that the rapid increase of $\dot P_{Shk}$ and $\dot T_{tr,\mu}$
with $a$ does not immediately imply that their actual
detection is facilitated as $P$ increases. Even though the timing signal 
increases with $P$, the number of transits, which determines the timing 
error, decreases as $P^{-1}$ for a given time interval over
which the system is being monitored. Using the results of Ford \etal (2008) 
and Heyl \& Gladman (2007) on transit timing precision we 
find that the time $\Pi_P$ one needs to monitor the transiting system to detect 
$\dot P$ induced by the proper motion scales as\footnote{This 
dependence is found by equating $\dot P_{Shk}\Pi_P$ to $\sigma_D/N$ where 
$N=\Pi_P/P$ is the number of observed transits and $\sigma_D$ is the 
uncertainty in measurement of $T_{tr}$, which is  
given by eq. (6) of Ford \etal (2008).} $\Pi_P\propto P^{4/15}$, i.e. it
increases with $P$ but not very rapidly: it takes $3.6$ times longer for 
$P=1$ yr system to get the same S/N for $\dot P_{Shk}$ due to proper motion as 
for the 3 d system. The uncertainty in $T_{tr}$ decreases with the
number of observed transits slower than the uncertainty in $\dot P$. As a 
result, the time $\Pi_{tr}$ one needs to monitor the transiting system to detect 
$\dot T_{tr}$ caused by the proper motion decreases with $P$ as\footnote{This 
follows from equating $\dot T_{tr,\mu}\Pi_{tr}$ to $\sigma_D$ 
given by eq. (6) of Ford 
\etal (2008).} 
$\Pi_{tr}\propto P^{-2/9}$. Thus, it is easier to measure 
$\dot T_{tr,\mu}$ in wide separation transiting systems.

Based on the results of Ford \etal (2008) it was estimated by Jord\'an
\& Bakos (2008) that the {\it Kepler} mission should be able to
achieve a timing precision of $\sim 1.5$ s in 1 yr observation of a 12th
magnitude Solar type star transited by a Jupiter-like planet
with $P=5$ d. One can easily deduce from this that a $3\sigma$ detection 
of $\dot T_{tr,\mu}=100$ ms yr$^{-1}$ (which is not unreasonable for proper 
motion) should take $\Pi_{tr}\approx 10$ yr of observations. 
Measurement error of $\dot P$
drops very rapidly with time but it would still take $\Pi_{P}\approx 70$ yr 
to achieve
a $3\sigma$ detection of $\dot P=100$ $\mu$s yr$^{-1}$ caused by the Shklovskii
effect. Thus, while one might hope to measure $\dot T_{tr,\mu}$ in some 
nearby, high-proper motion systems (like GJ436) on timescale of $\sim 10$ yr,
the measurement of $\dot P$ would likely require next generation facilities with 
photometric precision much higher than that of the {\it Kepler} mission.

In systems with low proper motion $\dot T_{tr,\mu}$ can be viewed as 
an irreducible 
systematic uncertainty to which quantities like $\dot T_{tr,GR}$ can be
measured. This is because even if $\mu$ is known precisely one still does not know 
a priori the angle $\beta$ (but see below) which determines $\dot T_{tr,\mu}$. 
Thus, proper motion limits to some extent our ability to interpret the 
measurement of 
$\dot T_{tr}$ in terms of the physical parameters of the system (e.g. $J_2$, 
etc.). Measurement of $\dot P$ does not suffer from this uncertainty 
since $\dot P_{Shk}$ is independent of $\beta$ and can thus be fully 
accounted for once $\mu$ is known from astrometric measurements.

Previous discussion has implicitly assumed that a 
transiting system moves at a constant 
speed with respect to observer. In reality observer is located
at Earth, which orbits the Sun. This to some extent 
complicates the analysis of the transit timing data since the relative velocity 
between the observed system and the Earth is a function of time with a 1 yr 
period. But given that the Earth-Sun motion is {\it known} 
its effect on transit timing can be easily accounted for: terrestrial  
orbital motion generates periodic apparent proper motion of any 
transiting system (even if it has no intrinsic proper motion) with respect 
to observer at Earth. The maximum amplitude of this apparent proper motion is 
$\mu_E\sim v_E/D\approx 60$ mas yr$^{-1}$, where $v_E\approx 30$ km s$^{-1}$. 
According to equations (\ref{eq:dTtrdt1}) and (\ref{eq:dTtrdt_num}) 
$\mu_E$ gives rise to periodically 
varying $\dot T_{tr,\mu}$ with an amplitude dependent on the orientation of the
orbital plane of the transiting system with respect to the ecliptic, potentially
providing a method of measuring angle $\beta$. The maximum possible value of such 
timing signal is about 
\ba
\dot T_{tr,E} \approx 30\frac{\mu}{100~\mbox{mas yr}^{-1}}
\frac{a/R_\star}{10}\frac{T_{tr}}{4~\mbox{hr}}~\mbox{ms yr}^{-1}.
\label{eq:Ttr_E}
\ea
It is also obvious that terrestrial orbital motion produces a periodic 
contribution to $\dot P$, even if the transiting system has zero intrinsic 
proper motion.

Such annual variations in $\dot T_{tr}$ and $\dot P$ can arise  
only as a result of the proper motion effects. One can hope to measure them
by properly combining the data on transit duration measured at different orbital
phases of the Earth. If such variations are 
detected then this periodic part of the timing signal can be used to constrain 
angle $\beta$ allowing one to remove the aforementioned systematic uncertainty in 
measuring other timing contributions.


\section{Conclusions.}
\label{sect:concl}

We investigated the effect of the proper motion on timing of transiting
exoplanetary systems in which the gravitational effect of other possible 
companions can be neglected. Proper motion re-orients planetary orbit with 
respect to our line of sight and changes the distance to the system.
Short period transiting systems having proper motion at the level of
$100$ mas yr$^{-1}$ should exhibit variation of the transit duration
at the level of $\sim 100$ ms yr$^{-1}$, which may be comparable to or 
exceed the timing signatures produced by the general relativity or 
stellar quadrupole and which should not be hard to detect. Proper motion
also causes variation of the observed orbital period through the Shklovskii
effect which dominates $\dot P$ for high proper motion systems.
Orbital motion of the Earth around the Sun gives rise to periodically
varying transit timing signal even in systems having zero intrinsic
proper motion.
Timing effects induced by the proper motion become especially
important in systems with zero eccentricity and in wide separation systems 
with periods longer than a month which should be discovered by {\it Kepler}.


\acknowledgements 

I am grateful to Ed Turner for 
useful discussions. 
This work made use of the data available through the SIMBAD 
Astronomical Database. 
The financial support for this work is provided
by the Sloan Foundation.


\begin{center}
\begin{deluxetable}{ l r r r r r r r r r r}
\tablewidth{0pc}
\tablecaption{Parameters of systems with transiting planets
\label{table}}
\tablehead{
\colhead{System}&
\colhead{$\mu_\alpha$}&
\colhead{$\mu_\delta$}&
\colhead{$D$}&
\colhead{$P$}&
\colhead{$a$}&
\colhead{$e$}&
\colhead{$\omega$}&
\colhead{$i$}&
\colhead{$R_\star$}&
\colhead{$M_\star$\\}
\colhead{}&
\colhead{(mas yr$^{-1}$)}&
\colhead{(mas yr$^{-1}$)}&
\colhead{(pc)}&
\colhead{(d)}&
\colhead{(AU)}&
\colhead{}&
\colhead{($^\circ$)}&
\colhead{($^\circ$)}&
\colhead{($R_\odot$)}&
\colhead{($M_\odot$)}
}
\startdata
HD149026 & -77.12 & 53.34 & $74.4\pm 7.2$ & 2.87588 & 0.0432 & 0 & - & 85.3 & 1.368 & 1.294 \\
HD189733 & -2.49 & -250.81 & $19.7\pm 1.0$ & 2.21857 & 0.0312 & 0 & - & 85.58 & 0.756 & 0.806 \\
HD17156 & 91 & -32.45 & $78.24$ & 21.21725 & 0.1594 & 0.6717 & 121.23 & 88.23 & 1.47 & 1.2 \\
GJ436 & 896.34 & -813.7 & $10.2\pm 0.2$ & 2.64385 & 0.02872 & 0.15 & 351 & 86.5 & 0.464 & 0.452 \\
TrES-4 & -8.1 & -33.0 & $485\pm 31$ & 3.55394 & 0.0488 & 0 & - & 82.81 & 1.816 & 1.394 \\
XO-5 & -32.4 & -24.4 & $270\pm 25$ & 4.187732 & 0.0508 & 0 & - & 86.8 & 1.11 & 1.0 \\
HAT-P-1 & 29.3 & -51.0 & $155\pm 15$ & 4.46529 & 0.0551 & 0.09 & 80.7 & 86.11 & 1.135 & 1.133 \\
WASP-2 & 3.0 & -53.1 & $157\pm 4$ & 2.152226 & 0.0307 & 0 & - & 84.81 & 0.84 & 0.89 \\
Sys-1 & 7.1 & 7.1 & $300$ & 100 & 0.4218 & 0.5 & 45 & 89.5 & 1.0 & 1.0 \\
Sys-2 & 7.1 & 7.1 & $300$ & 365 & 1.0 & 0.5 & 45 & 89.8 & 1.0 & 1.0  \\
\enddata
\end{deluxetable}
\end{center}

\begin{center}
\begin{deluxetable}{ l r r r r r r r r }
\tablewidth{0pc}
\tablecaption{Timing signatures in transiting systems
\label{table2}}
\tablehead{
\colhead{System}&
\colhead{$g$}&
\colhead{max $\dot\omega_\mu$}&
\colhead{$\dot\omega_{GR}$}&
\colhead{$\dot\omega_S$}&
\colhead{max $|\dot T_{tr,\mu}|$}&
\colhead{$\dot T_{tr,GR}$}&
\colhead{$\dot P_{Shk}$}&
\colhead{$\dot P_{\omega,GR}$\\}
\colhead{}&
\colhead{}&
\colhead{($10^{-7}$ yr$^{-1}$)}&
\colhead{($10^{-4}$ yr$^{-1}$)}&
\colhead{($10^{-7}$ yr$^{-1}$)}&
\colhead{(ms yr$^{-1}$)}&
\colhead{(ms yr$^{-1}$)}&
\colhead{($\mu$s yr$^{-1}$)}&
\colhead{($\mu$s yr$^{-1}$)}
}
\startdata
HD149026 & 5.5 & 4.5 & 7.1 & 114.8 & 24.0 & 0 & 12.46 & 0 \\
HD189733 & 11.4 & 12.2 & 7.8 & 23.4 & 69.3 & 0 & 18.2 & 0 \\
HD17156 & 6.2 & 4.7 & 0.44 & 1.8 & 21.2 & -97.2 & 102.6 & -2.4 \\
GJ436 & 31.9 & 58.7 & 4.1 & 3.6 & 616.6 & -201.1 & 261.8 & 13.85 \\
TrES-4 & 8.7 & 1.6 & 5.8 & 297.6 & 16.7 & 0 & 13.2 & 0 \\
XO-5 & 7.7 & 2.0 & 3.2 & 26.1 & 14.9 & 0 & 12.3 & 0 \\
HAT-P-1 & 11.5 & 2.9 & 3.2 & 20.6 & 24.8 & 9.8 & 15.9 & 1.7 \\
WASP-2 & 11.3 & 2.6 & 9.5 & 39.9 & 15.4 & 0 & 6.3 & 0 \\
Sys-1 & 49.2 & 0.48 & 0.02 & 0.04 & 30.7 & 6.7 & 19.9 & 0.3 \\
Sys-2 & 108 & 0.48 & $0.002$ & $0.002$ & 105.1 & $0.7$ & 72.5 & $0.05$ \\
\enddata
\end{deluxetable}
\end{center}

\end{document}